\documentclass[epj]{webofc}
\usepackage[varg]{txfonts}   

\newcommand{\Pom}{I\!\!P}
\newcommand{\Reg}{I\!\!R}

\newcommand{\bdPt}{\mbox{\boldmath $dP_{t}$}}
\newcommand{\bqta}{\mbox{\boldmath $q_{t,1}$}}
\newcommand{\bqtb}{\mbox{\boldmath $q_{t,2}$}}
\newcommand{\bpta}{\mbox{\boldmath $p_{t,1}$}}
\newcommand{\bptb}{\mbox{\boldmath $p_{t,2}$}}

\woctitle{MESON2018 - the 15$^\textrm{th}$ International Workshop on Meson Physics}
\begin{document}
\selectlanguage{english}
\title{Central exclusive production of $K^{+}K^{-}$ pairs in proton-proton collisions}

\author{Piotr Lebiedowicz\inst{1}\fnsep\thanks{\email{Piotr.Lebiedowicz@ifj.edu.pl}} \and
        Antoni Szczurek\inst{1,2}\and
        Otto Nachtmann\inst{3} 
}

\institute{Institute of Nuclear Physics Polish Academy of Sciences,
           ul. Radzikowskiego 152, PL 31-342 Krak{\'o}w, Poland
\and
           University of Rzesz\'ow, ul. Pigonia 1, PL 35-959 Rzesz{\'o}w, Poland
\and       Institut f\"ur Theoretische Physik, Universit\"at Heidelberg,
           Philosophenweg 16, D-69120 Heidelberg, Germany
          }

\abstract{
We discuss central exclusive diffractive production of light mesons 
in the reactions $pp \to pp\pi^{+}\pi^{-}$ and $pp \to ppK^{+}K^{-}$. 
The calculation is based on a tensor-pomeron approach. 
We include a purely diffractive dipion continuum, 
and the scalar $f_{0}(980)$, $f_{0}(1500)$, $f_{0}(1710)$ and tensor $f_{2}(1270)$, $f'_{2}(1525)$ resonances 
decaying into pseudoscalar meson pairs. 
We include also photoproduction mechanisms for the nonresonant (Drell-S\"oding) 
and the $\phi(1020)$ resonance contributions.
The theoretical results are compared with existing CDF experimental data 
and predictions for being carried out LHC experiments are presented. 
The distributions in dimeson invariant mass 
and in a special "glueball filter variable" 
including the interference effects of resonance and dimeson continuum are presented.
}
\maketitle
\section{Introduction}
\label{intro}

Diffractive studies are one of the important parts of the physics programme 
for the RHIC and LHC experiments. 
A particularly interesting classes are exclusive processes, where
all centrally produced particles are detected. 
An example is the process of kaon pair production
$pp \to ppK^{+}K^{-}$, which can be used for studies of low-mass resonances,
including searches for glueballs.
Such processes were studied extensively at CERN ISR \cite{ISR,ISR1,Breakstone:1989ty,Breakstone:1990at}
and WA102 \cite{WA102,WA102a} experiments
and are an attractive topic of current experimental studies: COMPASS \cite{COMPASS,Austregesilo:2016sss}, 
STAR \cite{Sikora:2016evz}, 
CDF \cite{Aaltonen:2015uva}, 
LHCb \cite{McNulty:2016sor,McNulty:2017ejl,Goncerz:2018xcw},
ALICE \cite{Schicker:2012nn},
CMS \cite{Khachatryan:2017xsi}.
Feasibility studies for the $p p \to p p \pi^+ \pi^-$ process with proton tagging
carried out for ATLAS + ALFA detectors are shown in \cite{Staszewski:2011bg}.

On the theoretical side, the main contribution to
the central diffractive exclusive production at high energies
can be understood as being due to the exchange of two pomerons/reggeons ($\Pom$/$\Reg$)
between the external nucleons and the centrally produced hadronic system.
One of the first calculations were concerned with continuum production
in the $p p \to p p \pi^+ \pi^-$ \cite{Lebiedowicz:2009pj,Lebiedowicz:2011nb,Staszewski:2011bg}, 
$p p \to nn \pi^+ \pi^+$ \cite{Lebiedowicz:2010yb},
and $p p \to p p K^+ K^-$ \cite{Lebiedowicz:2011tp} reactions,
where the amplitudes were written in terms of $\Pom/\Reg$ exchanges 
with parameters fixed from phenomenological analyses of
$NN$, $\pi N$ and $KN$ total and elastic scattering.

In more recent studies we describe the soft pomeron
as an effective rank-2 symmetric tensor exchange, 
as in the model of \cite{Ewerz:2013kda}.
In \cite{Ewerz:2016onn} it was shown that the tensor-pomeron model 
is consistent with the experimental data 
on the helicity structure of $pp$ elastic scattering at
$\sqrt{s} = 200$~GeV and small $|t|$ from the STAR experiment \cite{Adamczyk:2012kn}.
In \cite{Lebiedowicz:2013ika} the central exclusive production 
of several scalar and pseudoscalar mesons 
in the reaction $p p \to p p M$ was studied for the relatively low WA102 energy.
Then, the model was applied
to the $p p \to p p \pi^+ \pi^-$ \cite{Lebiedowicz:2016ioh}
and $p p \to p p K^+ K^-$ \cite{Lebiedowicz:2018eui} reactions at high energies
including the continuum and the dominant scalar and tensor resonances
decaying into the pseudoscalar meson pairs. 
In \cite{Lebiedowicz:2016ioh} we considered all (seven) possible tensorial structures 
for the $\Pom \Pom f_{2}$ coupling.
The resonant $\rho^0$ and non-resonant (Drell-S\"oding)
$\pi^{+}\pi^{-}$ photoproduction was studied in \cite{Lebiedowicz:2014bea}.
The $\rho^{0}$ production associated with 
a very forward/backward $\pi N$ system in the $pp \to p \rho^{0} (\pi N)$ processes
was studied in \cite{Lebiedowicz:2016ryp}.
The central exclusive $\pi^+ \pi^-\pi^+ \pi^-$ production 
via the intermediate $\sigma\sigma$ and $\rho^0\rho^0$ states in $pp$ collisions
was discussed in \cite{Lebiedowicz:2016zka}.
Recently, in \cite{Lebiedowicz:2018sdt}, the $pp \to pp p\bar{p}$ reaction was studied.
It should be emphasized that the latter process has a very different characteristics than
the $pp \to pp \pi^+ \pi^-$ or $pp \to pp K^+ K^-$ one,
which we predicted using the correct treatment of the spin degrees of freedom in
the calculations.

\section{Sketch of formalism}
\label{formalism}

\begin{figure}[ht]
\centering
(a)\includegraphics[width=4.0cm,clip]{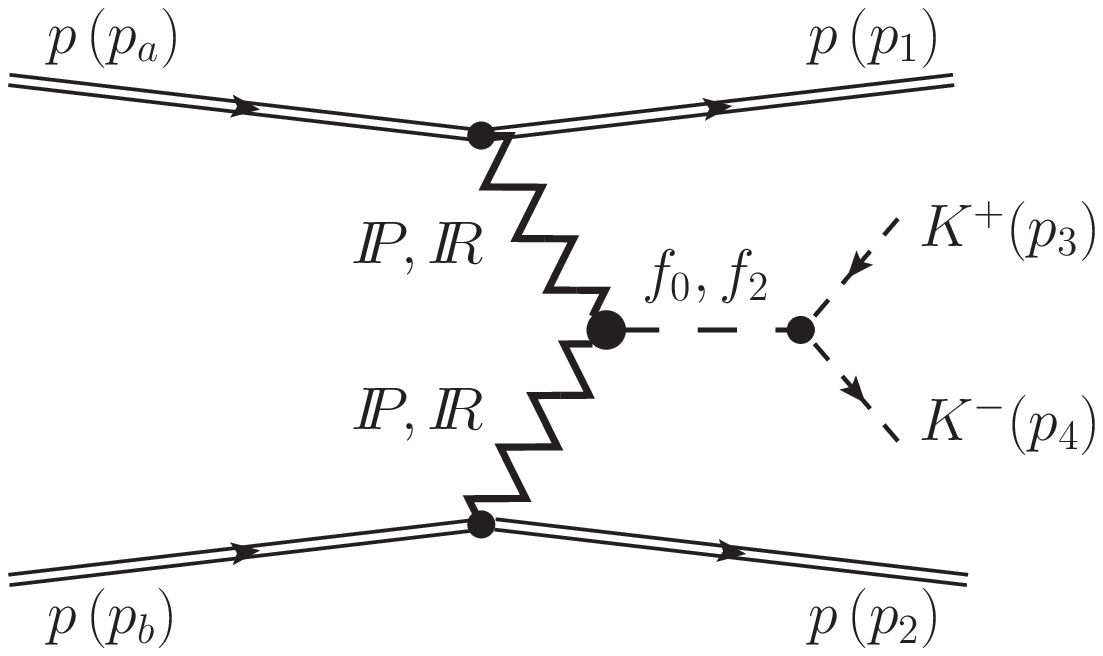}
(b)\includegraphics[width=3.4cm,clip]{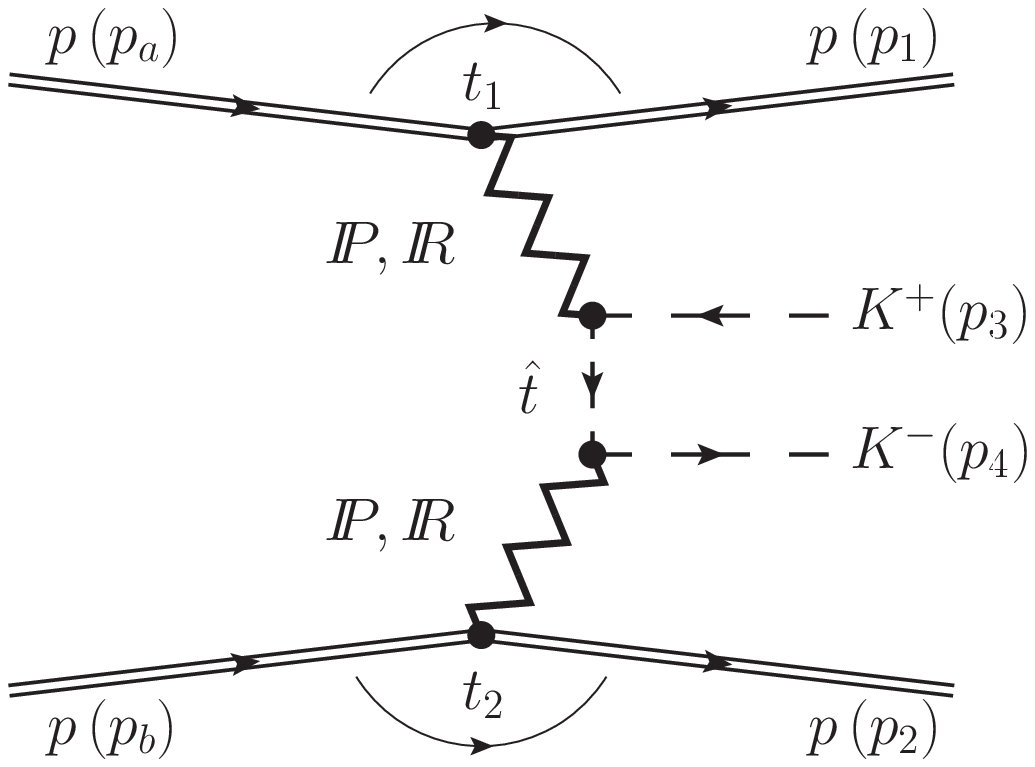}\\
(c)\includegraphics[width=3.6cm,clip]{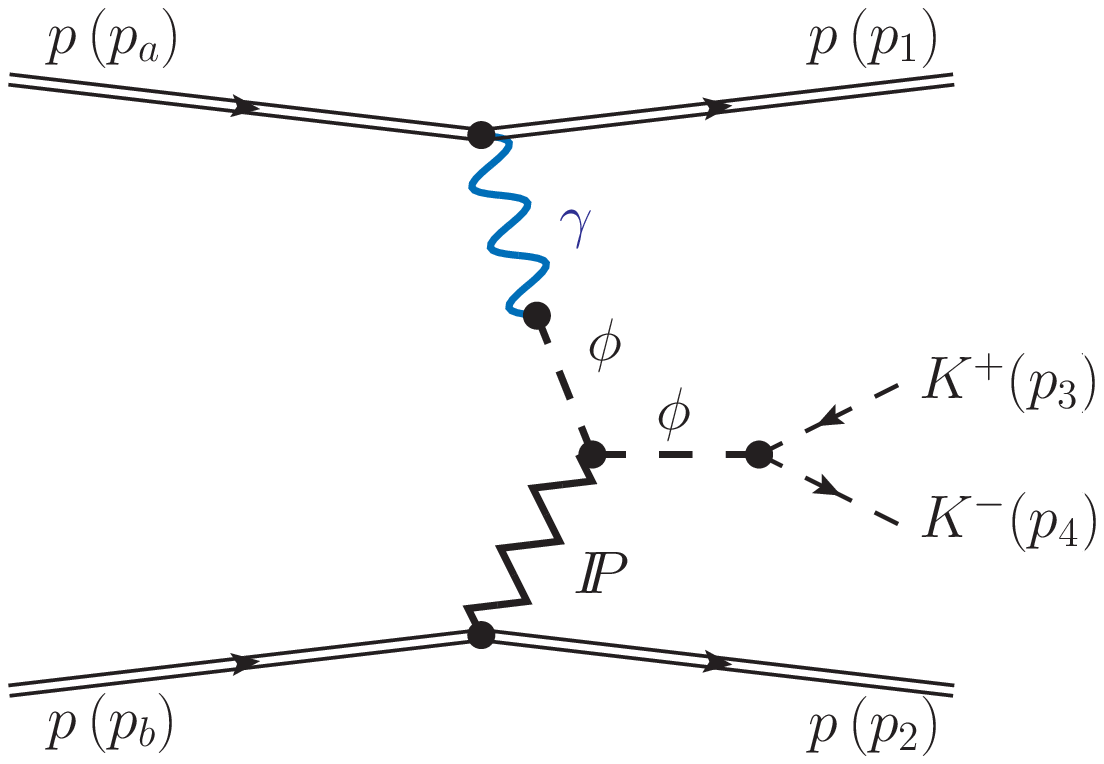}
(d)\includegraphics[width=3.4cm,clip]{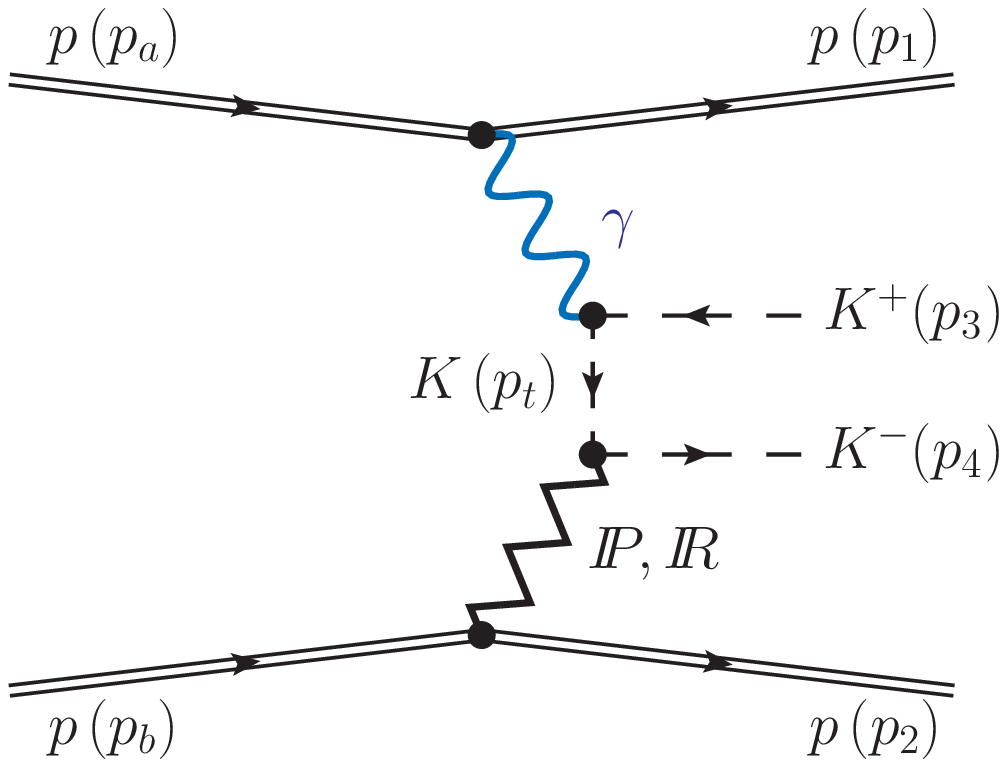}
   \includegraphics[width=3.4cm,clip]{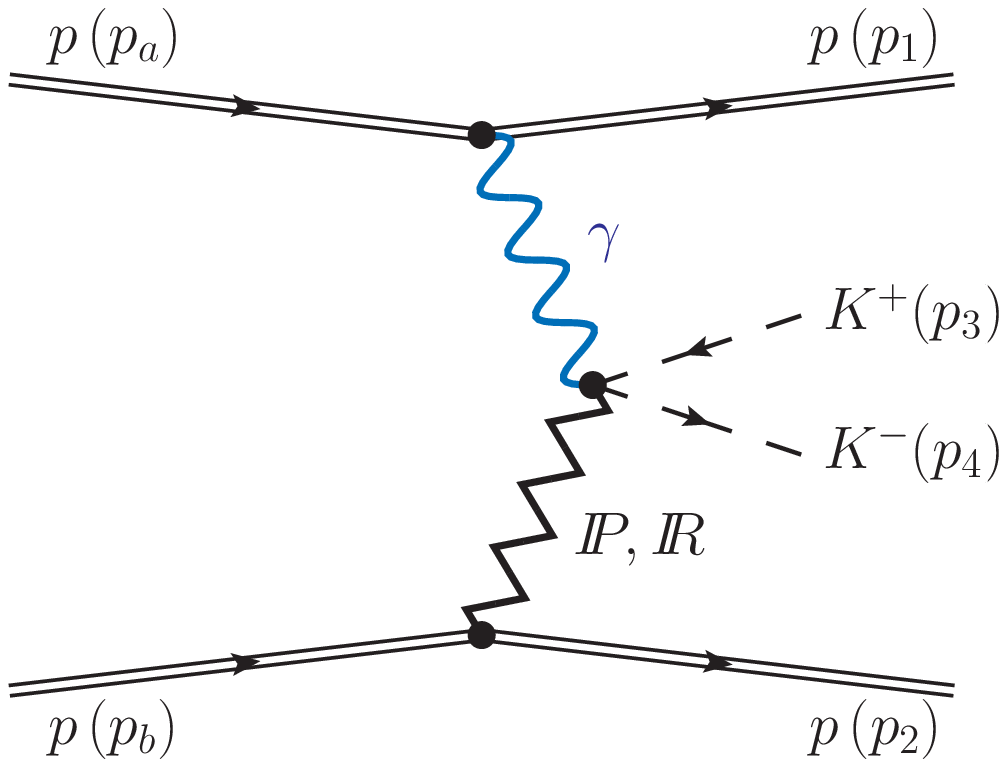}
\caption{
Schematic Born-level diagrams of: (a) central exclusive purely diffractive scalar and tensor resonance
and (b) continuum $K^+ K^-$ production, (c) $\phi$ resonance and (d) non-resonant photoproduction. 
}
\label{fig:0}
\end{figure}

The Born level diagrams for the $pp \to ppK^{+} K^{-}$ process
are shown in Fig.~\ref{fig:0}.
The total amplitude is a coherent sum of continuum amplitudes
and the amplitudes with the $s$-channel 
scalar ($f_{0}(980)$, $f_{0}(1500)$, $f_{0}(1710)$)
and tensor ($f_{2}(1270)$, $f'_{2}(1525)$) resonances.

For instance, the Born (without absorption effects) amplitude for the process
$pp \to pp (\Pom \Pom \to f_{2}'(1525) \to K^{+}K^{-})$ within the tensor-pomeron model can be written as
\begin{eqnarray}
&&{\cal M}^{(\Pom \Pom \to f_{2}'\to K^{+}K^{-})}_{\lambda_{a} \lambda_{b} \to \lambda_{1} \lambda_{2} K^{+}K^{-}} 
=  (-i)\,
\bar{u}(p_{1}, \lambda_{1}) 
i\Gamma^{(\Pom pp)}_{\mu_{1} \nu_{1}}(p_{1},p_{a}) 
u(p_{a}, \lambda_{a})\;
i\Delta^{(\Pom)\, \mu_{1} \nu_{1}, \alpha_{1} \beta_{1}}(s_{1},t_{1}) \nonumber \\
&& \qquad \qquad \qquad \qquad \times 
i\Gamma^{(\Pom \Pom f_{2}')}_{\alpha_{1} \beta_{1},\alpha_{2} \beta_{2}, \rho \sigma}(q_{1},q_{2}) \;
i\Delta^{(f_{2}')\,\rho \sigma, \alpha \beta}(p_{34})\;
i\Gamma^{(f_{2}' KK)}_{\alpha \beta}(p_{3},p_{4}) \nonumber \\
&& \qquad \qquad \qquad \qquad \times 
i\Delta^{(\Pom)\, \alpha_{2} \beta_{2}, \mu_{2} \nu_{2}}(s_{2},t_{2}) \;
\bar{u}(p_{2}, \lambda_{2}) 
i\Gamma^{(\Pom pp)}_{\mu_{2} \nu_{2}}(p_{2},p_{b}) 
u(p_{b}, \lambda_{b}) \,,
\label{amplitude_f2_pomTpomT}
\end{eqnarray}
where $t_{1} = q_{1}^{2} = (p_{1} - p_{a})^{2}$, $t_{2} = q_{2}^{2} = (p_{2} - p_{b})^{2}$, 
$s_{1} = (p_{a} + q_{2})^{2} = (p_{1} + p_{34})^{2}$,
$s_{2} = (p_{b} + q_{1})^{2} = (p_{2} + p_{34})^{2}$,
$p_{34} = p_{3} + p_{4}$. $\Delta^{(\Pom)}$ and $\Gamma^{(\Pom pp)}$ 
denote the effective tensor-pomeron propagator and proton vertex function, respectively.
For the explicit expressions see section~3 of \cite{Ewerz:2013kda}.
Other details as form of form factors, the tensor-meson propagator $\Delta^{(f_{2}')}$,
the $\Pom \Pom f_{2}'$ and $f_{2}' KK$ vertices are given in \cite{Lebiedowicz:2018eui}.
Our attempts to determine the parameters of pomeron-pomeron-meson couplings
as far as possible from experimental data have been presented in section~III of \cite{Lebiedowicz:2018eui}.
There are also the parameters of pomeron/reggeon-kaon couplings
obtained from fits to kaon-nucleon total cross section data.
The diffractive and photoproduction contributions to 
$K^{+}K^{-}$ production must be added coherently at the amplitude level 
and in principle could interfere.
In reality the Born approximation is not sufficient and 
absorption corrections (rescattering effects) have to be taken into account,
see e.g. \cite{Harland-Lang:2013dia,Lebiedowicz:2015eka}.

\section{Selected results}
\label{results}

\begin{figure}[h]
\centering
\includegraphics[width=6.4cm,clip]{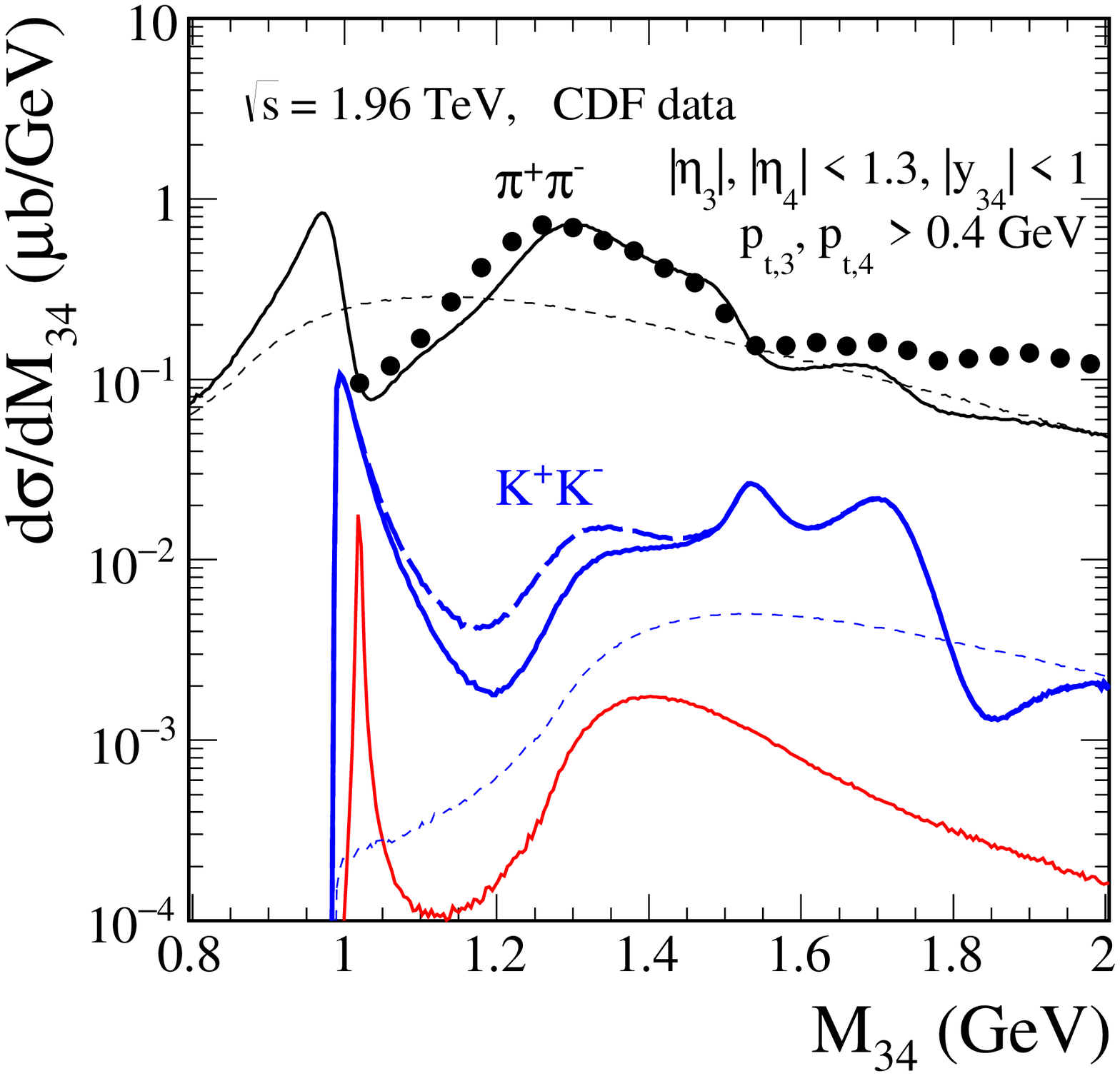}
\includegraphics[width=6.4cm,clip]{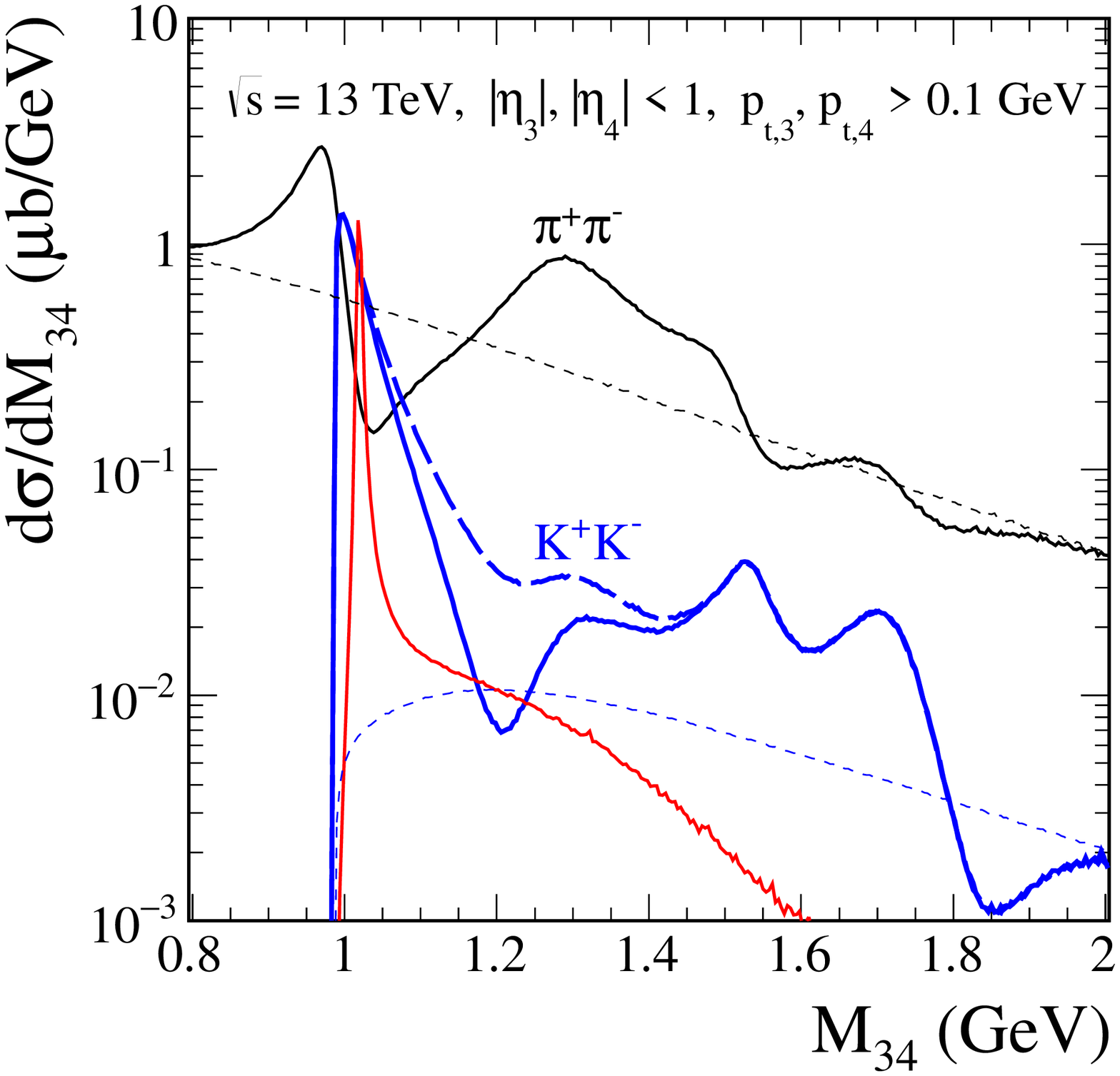}
\caption{The invariant mass distributions for
centrally produced $\pi^{+} \pi^{-}$ (the black top lines) 
and $K^{+} K^{-}$ (the blue bottom lines) pairs
with the relevant experimental kinematical cuts specified in the legend.
Results including both the non-resonant continuum and the resonances are presented.
The short-dashed lines represent the purely diffractive continuum term.
For the $pp \to ppK^{+}K^{-}$ reaction the solid and long-dashed blue lines correspond to 
the results for $\phi_{f_{0}(980)} = 0$ and $\pi/2$ in the coupling 
constant $g_{f_{0}(980) K^{+} K^{-}} \,e^{i \phi_{f_{0}(980)}}$, respectively.
The lower red line represents the $\phi(1020)$ meson 
plus continuum photoproduction contribution.
The CDF experimental data from \cite{Aaltonen:2015uva} in the top left panel
for the $p\bar{p} \to p\bar{p} \pi^{+} \pi^{-}$ reaction are shown for comparison.
Absorption effects were taken into account effectively by the gap survival factors,
$\langle S^{2} \rangle = 0.1$ for the purely diffractive contributions
and $\langle S^{2} \rangle = 0.9$ for the photoproduction contributions.}
\label{fig:dsig_dM34}
\end{figure}

In Fig.~\ref{fig:dsig_dM34} we present the dikaon invariant mass distributions 
(the blue and red lines) imposing experimental cuts.
We also show, for comparison, the purely diffractive contribution
for the central production of $\pi^{+} \pi^{-}$ pairs (see the upper black lines).
The short-dashed lines represent the purely diffractive continuum term 
including both pomeron and reggeon exchanges,
discussed in \cite{Lebiedowicz:2016ioh,Lebiedowicz:2018eui}.
For the $pp \to ppK^{+}K^{-}$ reaction 
the solid lines represent the coherent sum of the diffractive
continuum, and the scalar $f_{0}(980)$, $f_{0}(1500)$, $f_{0}(1710)$,
and tensor $f_{2}(1270)$, $f'_{2}(1525)$ resonances.
The lower red lines show the photoproduction term including
the dominant $\phi(1020) \to K^{+}K^{-}$ and the continuum (Drell-S\"oding) contributions.
The narrow $\phi$ resonance is visible above the continuum term.
It may in principle be visible on top of the broader $f_{0}(980)$ resonance.
The coupling parameters of the tensor pomeron to the $\phi$ meson have been
fixed based on the HERA experimental data for the $\gamma p \to \phi p$ reaction \cite{HERA,HERA1}.

\begin{figure}[h]
\centering
(a)\includegraphics[width=6cm,clip]{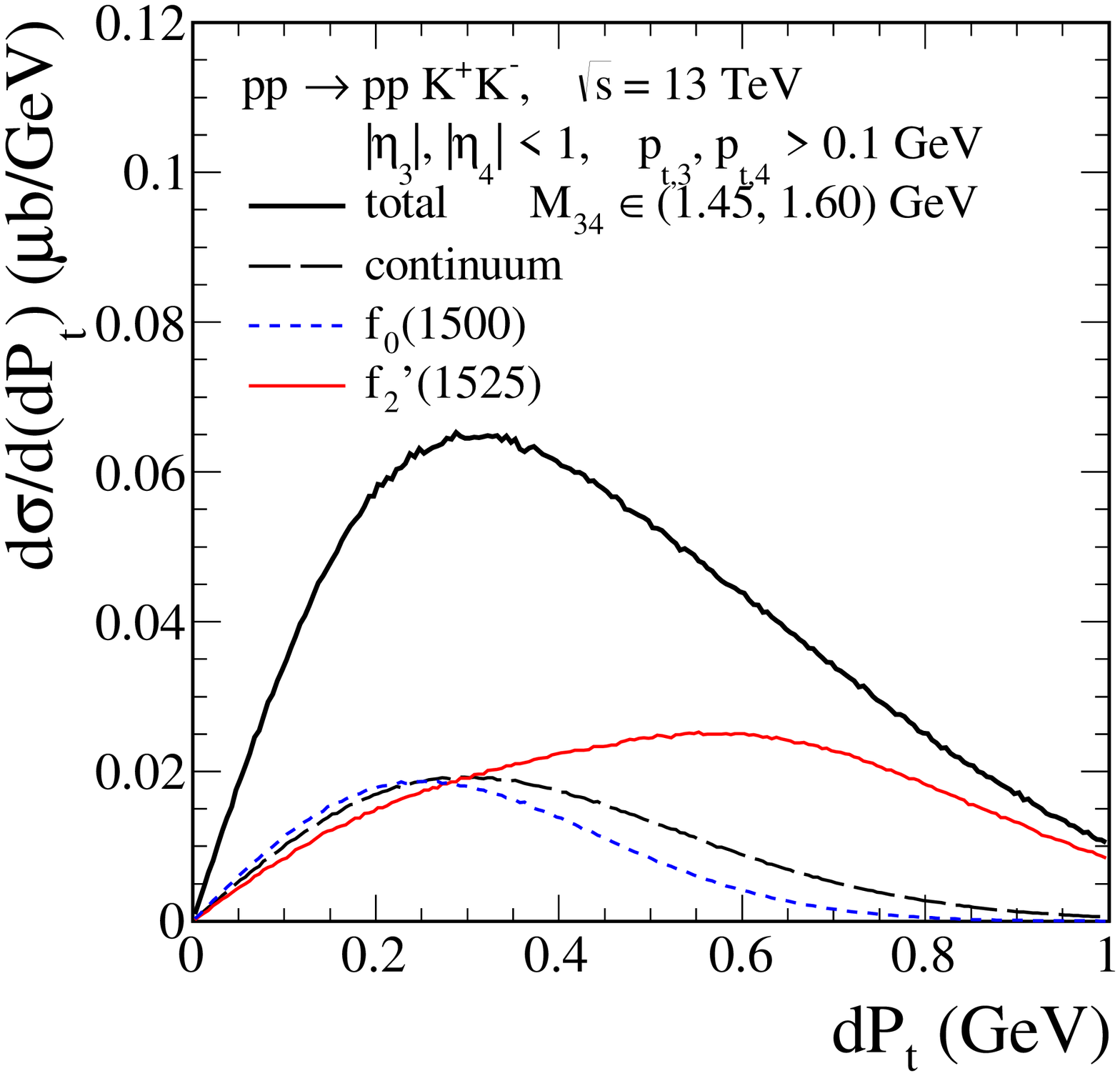}
(b)\includegraphics[width=6cm,clip]{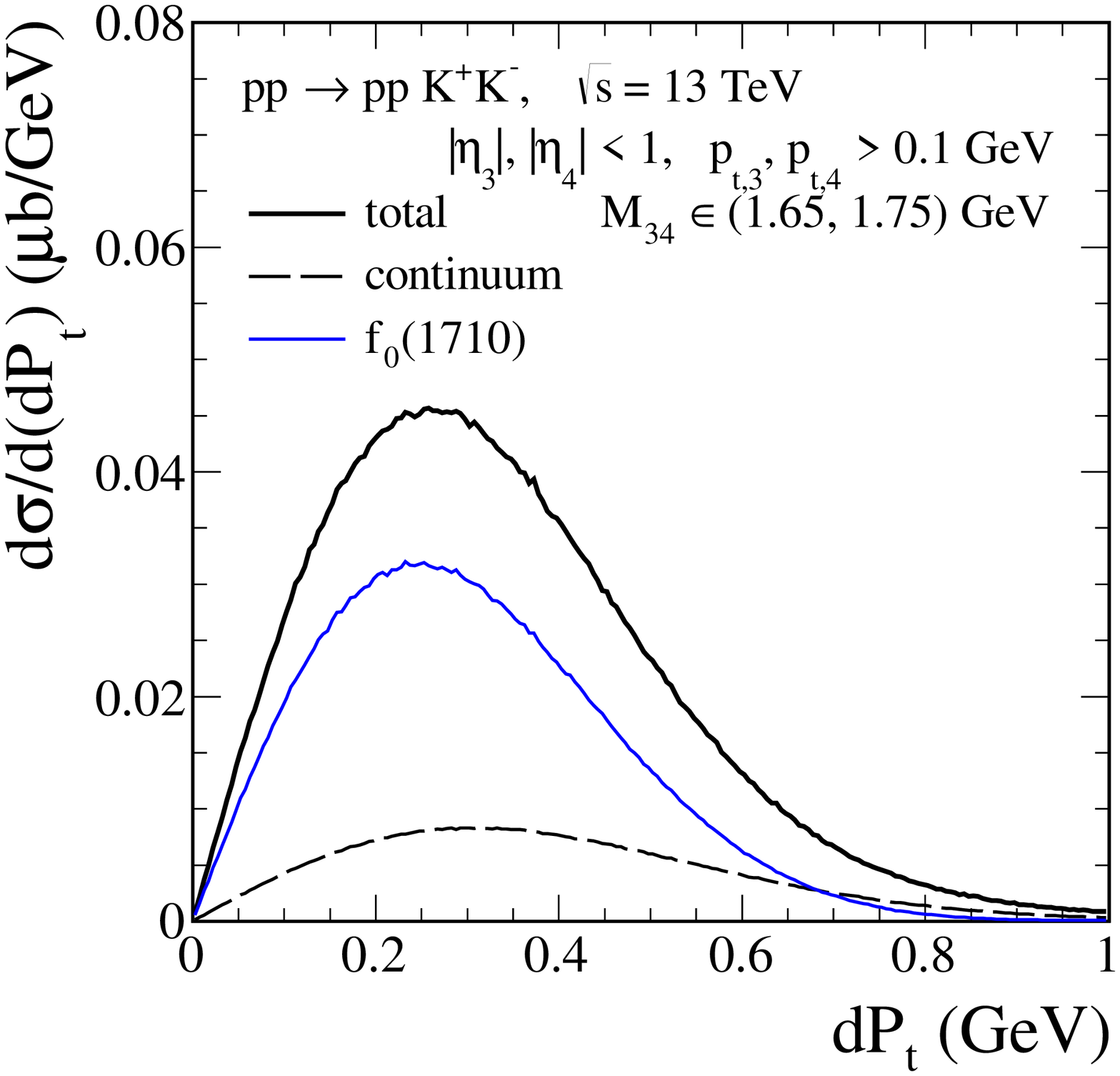}
\caption{
The differential cross sections $d\sigma/d(dP_{t})$
as a function of the $dP_{t}$ ``glueball filter'' variable
for the $pp \to pp K^{+} K^{-}$ reaction.
Calculations were done for $\sqrt{s} = 13$~TeV, $|\eta_{K}| < 1$,
$p_{t,K} > 0.1$~GeV, and in two dikaon invariant mass regions:
(a) $M_{34} \in (1.45, 1.60)$~GeV and (b) $M_{34} \in (1.65, 1.75)$~GeV.
No absorption effects were taken into account here.}
\label{fig:dsig_dptperp}
\end{figure}

In Fig.~\ref{fig:dsig_dptperp} we present distributions 
in a special "glueball filter variable" $dP_{t}$ \cite{Close:1997pj}
defined by the difference of the transverse momentum vectors $dP_{t} = |\bdPt|$,
$\bdPt = \bqta - \bqtb = \bptb - \bpta$.
Results for the ALICE kinematics and for two regions of:
(a) $M_{34} \in (1.45, 1.60)$~GeV and 
(b) $M_{34} \in (1.65, 1.75)$~GeV are shown.
We see that the maximum for the $q \bar{q}$ state $f'_{2}(1525)$
is around of $dP_{t} = 0.6$~GeV. On the other hand,
for the scalar glueball candidates $f_{0}(1500)$ and $f_{0}(1710)$
the maximum is around $dP_{t} = 0.25$~GeV, that is,
at a lower value than for the $f'_{2}(1525)$.
This is in accord with the discussion in Refs.~\cite{WA102,WA102a}.

\section{Conclusions}
\label{conclusions}

In our recent paper \cite{Lebiedowicz:2018sdt} we have analysed 
the central exclusive production (CEP) of $K^{+}K^{-}$ pairs 
in proton-proton collisions at high energies. 
We have taken into account purely diffractive and diffractive photoproduction mechanisms.
For the purely diffractive mechanism we have included
the continuum and the dominant scalar $f_{0}(980)$, $f_{0}(1500)$, $f_{0}(1710)$
and tensor $f_{2}(1270)$, $f'_{2}(1525)$ resonances decaying
into $K^{+} K^{-}$ pairs.
The amplitudes have been calculated using Feynman rules within
the tensor-pomeron model \cite{Ewerz:2013kda}.
The effective Lagrangians and the vertices for $\Pom \Pom$ fusion 
into the scalar and tensor mesons were discussed in \cite{Lebiedowicz:2013ika}
and \cite{Lebiedowicz:2016ioh}, respectively.
Some model parameters (pomeron-pomeron-meson couplings, 
the off-shell dependence of form factors $\Lambda_{off,M} = 0.7$~GeV)
have been roughly adjusted to CDF data \cite{Aaltonen:2015uva}
and then used for predictions for the STAR, ALICE, ATLAS, CMS and LHCb experiments.

Exclusive production of light mesons both in the $pp \to pp\pi^{+}\pi^{-}$ and $pp \to ppK^{+}K^{-}$ reactions
are measurable at RHIC and LHC.
We have focused mainly on the invariant mass distributions of centrally produced $K^{+}K^{-}$.
The pattern of visible structures in the invariant mass distributions
is related to the scalar and tensor isoscalar mesons
and it depends on experimental kinematics.
One can expect, with our default choice of parameters,
that the scalar $f_{0}(980)$, $f_{0}(1500)$, $f_{0}(1710)$
and the tensor $f_{2}(1270)$, $f'_{2}(1525)$ mesons
will be easily identified experimentally in CEP.
The distributions, in the so-called glueball filter variable $dP_{t}$, show different behavior
in the $K^{+}K^{-}$ invariant mass windows 
around glueball candidates with masses $\sim 1.5$~GeV and $\sim 1.7$~GeV
than in other regions.

The absorption effects lead to a huge damping of the cross section for the purely
diffractive contribution and a relatively small reduction
of the cross section for the $\phi(1020)$ photoproduction contribution.
Therefore we expect that that one could observe the $\phi$ resonance term,
especially when no restrictions on the leading protons are included.
Finally we note that central exclusive production of $\phi$ 
offers also the possibility to search for
effects of the elusive odderon, as was pointed out in \cite{Schafer:1991na}.
A Monte Carlo generator containing a various processes 
and including detector effects (acceptance, efficiency)
would be useful in theory-data comparison.
The $\mathtt{GenEx}$ Monte Carlo generator \cite{GENEX,Kycia:2017ota} 
could be used in this context.

\begin{acknowledgement}
This work was partially supported by Polish Ministry of Science and Higher Education 
under the Iuventus Plus grant (IP2014~025173)
and by Polish National Science Centre under the grants DEC-2014/15/B/ST2/02528 and DEC-2015/17/D/ST2/03530.
\end{acknowledgement}

%
%
%

\end{document}